\documentclass[conference]{IEEEtran}
\IEEEoverridecommandlockouts

\usepackage{cite}
\usepackage{todonotes}
\usepackage{amsmath,amssymb,amsfonts}
\usepackage{algorithmic}
\usepackage{graphicx}
\usepackage{textcomp}
\usepackage[binary-units=true]{siunitx}
\usepackage{subcaption}
\usepackage{xcolor}
\usepackage{tikz}
\usetikzlibrary{patterns}
\usepackage{pgfplots}
\usepackage{pgfplotstable}
\usepgfplotslibrary{groupplots}
\pgfplotsset{compat=1.18}

\usepackage{flushend}

\usepackage[hidelinks]{hyperref}

\def\BibTeX{{\rm B\kern-.05em{\sc i\kern-.025em b}\kern-.08em
    T\kern-.1667em\lower.7ex\hbox{E}\kern-.125emX}}
\begin{document}

\title{A Tabular Schedule Abstraction for Communication-Aware Evaluation of Pipeline-Parallel LLM Training\\
\thanks{This work is part of the "Model-Based AI" project, which is funded by the Carl Zeiss Foundation. The authors acknowledge support by the state of Baden-Württemberg through bwHPC
and the German Research Foundation (DFG) through grant INST 35/1597-1 FUGG.}
}

\author{
\IEEEauthorblockN{
Daniel Barley\IEEEauthorrefmark{1},
Jonathan Leis\IEEEauthorrefmark{1},
Benjamin Klenk\IEEEauthorrefmark{2},
Holger Fr\"oning\IEEEauthorrefmark{1}
}
\IEEEauthorblockA{\IEEEauthorrefmark{1}
Hardware and Artificial Intelligence (HAWAII) Lab, Heidelberg University, Heidelberg, Germany}
\IEEEauthorblockA{\IEEEauthorrefmark{2}
NVIDIA Corporation, Santa Clara, CA, USA}
\IEEEauthorblockA{
\{daniel.barley, holger.froening\}@ziti.uni-heidelberg.de,
jonathan.leis@stud.uni-heidelberg.de,
bklenk@nvidia.com
}
}

\maketitle

\begin{abstract}
Pipeline parallelism is a key technique for distributed training of large
language models because it reduces per-device parameter and activation memory.
However, comparing pipeline schedules is difficult: analytical models expose
structural quantities such as bubble ratios, while end-to-end hardware
experiments are costly and system-specific. In this work, we introduce a
tabular schedule abstraction and a unified multi-abstraction methodology that
connects formula-based reasoning, idealized schedule tables, and
communication-aware execution simulation.

Using this framework, we compare GPipe, 1F1B, Chimera, and Hanayo in its
restricted regime across multiple modeled system configurations. Our results
show that schedule rankings are not abstraction-invariant: communication can
negate structural advantages suggested by bubble analysis alone. Under the
assumptions considered here, GPipe and 1F1B are runtime-equivalent, but 1F1B
achieves a lower activation-memory peak. Chimera is advantageous mainly at low
microbatch counts and in communication-favorable regimes, while Hanayo is
effective in its intended restricted operating point but remains sensitive to
network bottlenecks. We further study an asymmetric Chimera-style placement,
which does not reduce the global peak memory requirement but reveals limited
runtime gains in shallow pipelines. Overall, pipeline schedule quality is
meaningful only in the context of the modeled execution environment.
\end{abstract}

\section{Introduction}\label{sec:introduction}

The training of large language models (LLMs) has emerged as a central
large-scale distributed systems problem. As model size, sequence length, and
training-token budgets continue to increase, single-device training is no
longer feasible for practically relevant regimes
\cite{brown2020language,kaplan2020scaling}.
Contemporary training stacks
therefore combine several forms of parallelism, most prominently data
parallelism, tensor parallelism, and pipeline parallelism, in order to
distribute both model state and computation across many accelerators
\cite{shoeybi2019megatron,narayanan2021efficient,rajbhandari2020zero}.

While this distribution enables the training of models at unprecedented scale,
it also fundamentally changes the performance bottlenecks that determine
overall efficiency. Training throughput is no longer governed solely by local
compute capability and memory bandwidth, but increasingly by inter-device
communication, synchronization, load imbalance, and schedule-induced idle time.

Within this design space, pipeline parallelism occupies a particularly
important role because it reduces the per-device parameter and activation
footprint by partitioning the model along network depth. However, the
effectiveness of pipeline-parallel training depends not only on how layers are
assigned to stages, but also on how forward and backward computations are
scheduled across those stages
\cite{huang2019gpipe,narayanan2019pipedream,narayanan2021memory,li2021chimera,liu2023hanayo}.
Even for an identical partitioning of model parameters,
different pipeline schedules can exhibit substantially different runtime and
memory characteristics. These differences arise from variations in pipeline
fill and drain behavior, activation retention time, communication exposure, and
the degree to which communication can be overlapped with useful computation. As
a result, pipeline scheduling should not be viewed merely as a theoretical
question of bubble minimization, but as a systems problem shaped by the
interaction of computation, memory, and communication under concrete execution
constraints.

Despite its importance, the comparative evaluation of pipeline schedules
remains methodologically challenging. Existing analyses often lie at one of two
extremes. On the one hand, analytical or structurally idealized treatments can
provide useful insight into quantities such as bubble ratios, memory peaks, or
asymptotic utilization, but necessarily abstract away communication costs,
cross-stage dependencies, and overlap effects. On the other hand, end-to-end
experiments on large hardware installations provide realistic measurements, yet
they are costly, difficult to control, and often tied to a particular system
configuration, making systematic cross-regime comparison difficult. Between
these extremes, simulation-based approaches provide an attractive middle ground
by preserving explicit representations of computation, communication, and
dependencies under configurable system assumptions
\cite{jia2019beyond,won2023astra}.

What is missing is a comparative workflow that connects these abstraction levels:
one that preserves the interpretability of structural analysis while still
incorporating the communication- and dependency-aware effects that ultimately
shape distributed execution.

This work addresses that gap through a simulation-based study of synchronous
pipeline schedules for distributed LLM training. Our work builds on Graphculon,
a graph-based execution modeling and simulation framework for distributed LLM
workloads, and extends it with a generalized schedule abstraction that
represents pipeline schedules in a uniform tabular form. This abstraction
separates schedule policy from execution-graph construction, simplifies the
implementation of new schedules, and enables systematic extraction of
structural metrics such as idle slots, fill/drain behavior, and activation
lifetimes. In addition, it provides the basis for translating schedule
descriptions into communication-aware execution graphs that can be simulated
under configurable system assumptions. This enables a controlled comparison
between three complementary views of a schedule: formula-based reasoning,
idealized schedule tables, and runtime-oriented execution simulation.

Using this extended framework, we compare representative synchronous pipeline
schedules, including GPipe~\cite{huang2019gpipe}, 1F1B~\cite{narayanan2019pipedream,narayanan2021memory}, Chimera~\cite{li2021chimera}, and Hanayo~\cite{liu2023hanayo} (in its restricted regime),
across multiple system regimes. Our results show that schedule rankings are
generally not abstraction-invariant: conclusions drawn from formulas or
idealized schedule tables do not necessarily carry over once communication and
dependency structure are modeled explicitly. Under the assumptions considered
in this work, GPipe and 1F1B are runtime-equivalent, but 1F1B achieves a lower
peak memory footprint. Chimera can outperform unidirectional schedules, but
primarily in regimes with small microbatch counts where its bidirectional
execution can reduce structural idle time without being dominated by additional
communication overhead. Hanayo is effective in its intended operating regime,
yet remains sensitive to communication bottlenecks. These findings suggest that
schedule quality should be interpreted as system-dependent rather than
universal, and that structurally attractive schedules may lose their advantage
when evaluated under more explicit execution models.

Beyond this comparative analysis, we also investigate a novel asymmetric
Chimera-style placement in which model stages are distributed unevenly across
the two counter-propagating branches. This case study is motivated by the
hypothesis that asymmetric placement may alleviate memory pressure or improve
runtime by reshaping where the critical path arises. The study indicates that
asymmetric placements may yield limited runtime benefits in specific
shallow-pipeline and communication-favorable regimes. However, our results do
not support the anticipated memory advantage: the evaluated asymmetric 1:2
placement does not reduce the relevant per-device memory peak. This refines the
design intuition around asymmetric schedules and illustrates the need for
explicit bottleneck-aware evaluation.

Overall, this paper makes three contributions.
\begin{itemize}
	\item First, we introduce a generalized tabular schedule abstraction for
	Graphculon that improves the expressiveness and comparability of pipeline
	schedule modeling.
	\item Second, we establish a unified multi-abstraction methodology for
	evaluating pipeline schedules, spanning formulas, structural schedule
	representations, and communication-aware runtime simulation.
	\item Third, we use this methodology to derive comparative insights into
	established schedule families and to assess a novel asymmetric Chimera
	variant.
\end{itemize}

Taken together, our results argue for a more cautious and methodologically
layered interpretation of pipeline schedule quality: efficient schedule design
cannot be inferred reliably from structural formulas alone, but must be
evaluated in the presence of communication, dependency, and memory
interactions.
 \section{Background}\label{sec:background}
Training large language models in distributed environments requires not only
partitioning model state and computation across many devices, but also
coordinating when intermediate results are produced, communicated, and
consumed. In this context, pipeline parallelism is attractive because it
reduces per-device model-state and activation requirements by partitioning the
model along network depth and executing different microbatches concurrently
across pipeline stages \cite{huang2019gpipe,narayanan2021efficient}. However,
its efficiency is governed not only by the stage partitioning itself, but also
by the schedule that determines how forward and backward computations progress
through the pipeline. This section briefly reviews the schedule families
considered in this work and highlights why their comparison is not trivial.

\subsection{Pipeline-parallel training and scheduling}
\label{sec:schedules}
In pipeline-parallel training, a minibatch is divided into multiple
microbatches that are injected into a sequence of pipeline stages. This enables
different stages to process different microbatches concurrently, thereby
improving device utilization compared to purely sequential execution
\cite{huang2019gpipe}. The schedule determines in which order microbatches and
their associated forward and backward phases are executed on each stage, and it
therefore directly affects runtime, idle time, and memory usage.

The baseline is \emph{GPipe}~\cite{huang2019gpipe}, which follows a fill-drain
strategy: forward passes are first injected until the pipeline is full, and
backward computation starts only after the full minibatch has reached the last
stage (Figure~\ref{fig:schedule_abstraction}a). 
This schedule is simple and exposes a clear structural bubble, but it retains activations for many microbatches
simultaneously and thus incurs high activation-memory pressure. A more memory-
efficient alternative is \emph{1F1B} (one-forward-one-backward, Figure~\ref{fig:schedule_abstraction}b), introduced in
PipeDream and later adapted to synchronous execution in PipeDream-Flush, where
stages alternate between forward and backward work once steady state has been
reached \cite{narayanan2019pipedream,narayanan2021memory}. This reduces
activation-retention intervals while largely preserving the pipeline structure.

More recent schedules modify the flow of work more aggressively.
\emph{Chimera} (Figure~\ref{fig:schedule_abstraction}c) uses two counter-propagating pipelines to reduce structural
bubbles through bidirectional execution, at the cost of increased communication
and duplicated parameters \cite{li2021chimera}. \emph{Hanayo} introduces a
wave-like synchronous schedule (Figure~\ref{fig:schedule_abstraction}d) that aims to improve utilization without
requiring the same degree of parameter duplication \cite{liu2023hanayo}. These
schedules illustrate that pipeline scheduling is not a one-dimensional
optimization problem: reducing bubble ratio may require additional
communication, different parameter placement, or longer dependency chains.

Two structural quantities are especially important in this context. The first is
the \emph{pipeline bubble}, i.e., the fraction of time during which workers are
idle because the pipeline is still filling, draining, or blocked by
dependencies. The second is \emph{activation lifetime}, which determines how
long forward activations must remain resident until the corresponding backward
computation consumes them. Bubble behavior primarily affects utilization and
runtime, whereas activation lifetime strongly affects peak memory consumption.
Both are schedule-dependent and therefore central to any meaningful comparison
of pipeline schedules.

At the same time, these structural quantities do not fully determine practical
performance. A schedule with an attractive bubble ratio may still perform worse
if it exposes more communication, if communication cannot be overlapped with
useful computation, or if additional persistent memory terms dominate the
activation savings. This motivates the distinction between \emph{structural
reasoning}, which studies schedules through formulas or idealized slot-based
representations, and \emph{communication-aware execution modeling}, which
incorporates explicit compute, communication, and dependency effects under
system assumptions.

\subsection{Difficulty of schedule comparison}

The comparative evaluation of pipeline schedules is difficult because different
evaluation methods expose different aspects of schedule behavior. Analytical
models provide compact expressions for quantities such as bubble ratios,
nominal utilization, or peak activation counts, and are therefore useful for
developing structural intuition \cite{huang2019gpipe,li2021chimera}. However,
they abstract away many effects that are central in distributed execution,
including communication cost, overlap constraints, and dependency-induced
serialization. As a result, analytically favorable schedules need not remain
favorable once those effects are taken into account.

At the other extreme, end-to-end experiments on hardware provide realistic
measurements, but they are costly to run, difficult to control, and often tied
to a particular system and software stack \cite{narayanan2021efficient,li2021chimera,liu2023hanayo}.
This makes it difficult to isolate schedule-specific behavior from other
confounding factors and limits systematic exploration across system regimes.
Simulation-based approaches provide a useful middle ground by representing
computation, communication, and dependencies explicitly while keeping the
underlying system parameters configurable \cite{jia2019beyond,won2023astra}.

The key challenge is therefore not simply to measure a schedule under one
particular setup, but to determine whether its qualitative ranking remains
stable across abstraction levels and system regimes. 
 \section{A Multi-Abstraction Evaluation Framework }
\label{sec:method}

This section introduces the schedule representation, its translation into an
execution model, and the abstraction levels and metrics used in the subsequent
evaluation.

\subsection{Tabular schedule abstraction}
\label{sec:tabular_schedule_abstraction}

We represent a pipeline schedule as a discrete table
over workers and time steps. Let $W$ denote the number of workers and $T$ the
number of schedule slots. A schedule is defined as
\[
S \in \left( \mathcal{M} \times \mathcal{P} \;\cup\; \{\emptyset\} \right)^{W
\times T},
\]
where $\mathcal{M}$ is the set of microbatches, $\mathcal{P}$ the set of
execution phases, and $\emptyset$ denotes an idle slot. Thus, each entry
$S_{w,t}$ specifies that worker $w$ executes phase $p \in \mathcal{P}$ for
microbatch $m \in \mathcal{M}$ at slot $t$, or remains idle otherwise. In the
setting considered here, $\mathcal{P} =
\{\texttt{fwd}, \texttt{agrad}, \texttt{wgrad}, \texttt{opt},
\texttt{recomp}\}$, corresponding to forward computation, activation-gradient
computation, weight-gradient computation, optimizer update, and optional
activation recomputation.

This representation is intentionally structural rather than temporal. A table
entry indicates that a phase is scheduled in a given slot, but does not yet
assign a hardware-dependent duration to that slot. The abstraction therefore
separates \emph{schedule policy} from \emph{execution cost}. This is useful for
two reasons. First, it makes schedule properties such as idle slots, fill and
drain behavior, and activation-retention intervals directly visible. Second,
it allows different schedule families to share a common representation, while
their communication and runtime implications are introduced only later through
execution-graph construction and simulation.

\begin{figure}
	\centering
	\begin{subfigure}{\linewidth}
		\includegraphics[width=\linewidth]{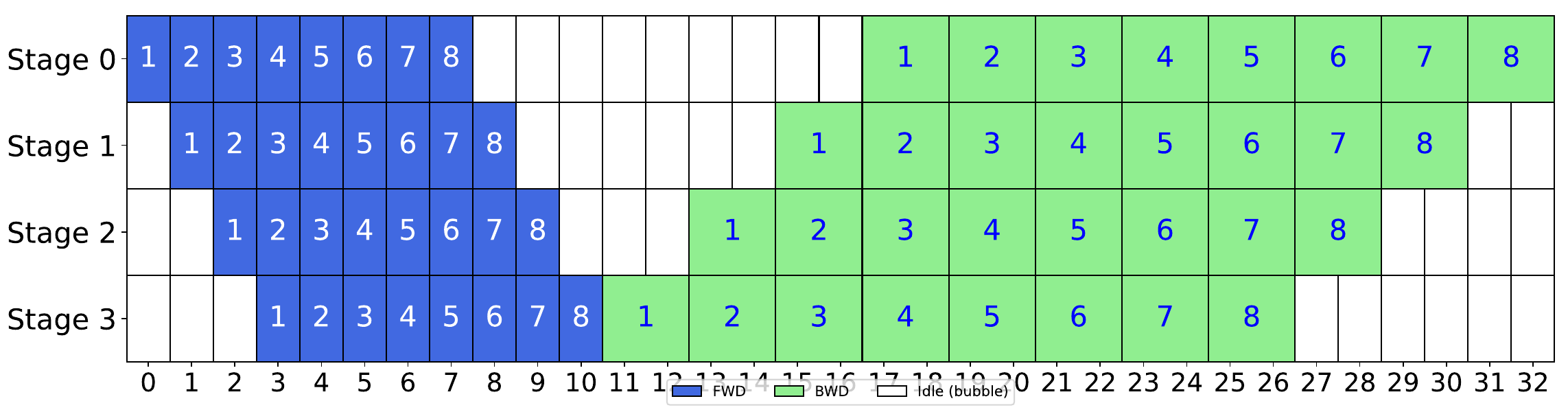}
		\caption{GPipe}
	\end{subfigure}
	\begin{subfigure}{\linewidth}
		\includegraphics[width=\linewidth]{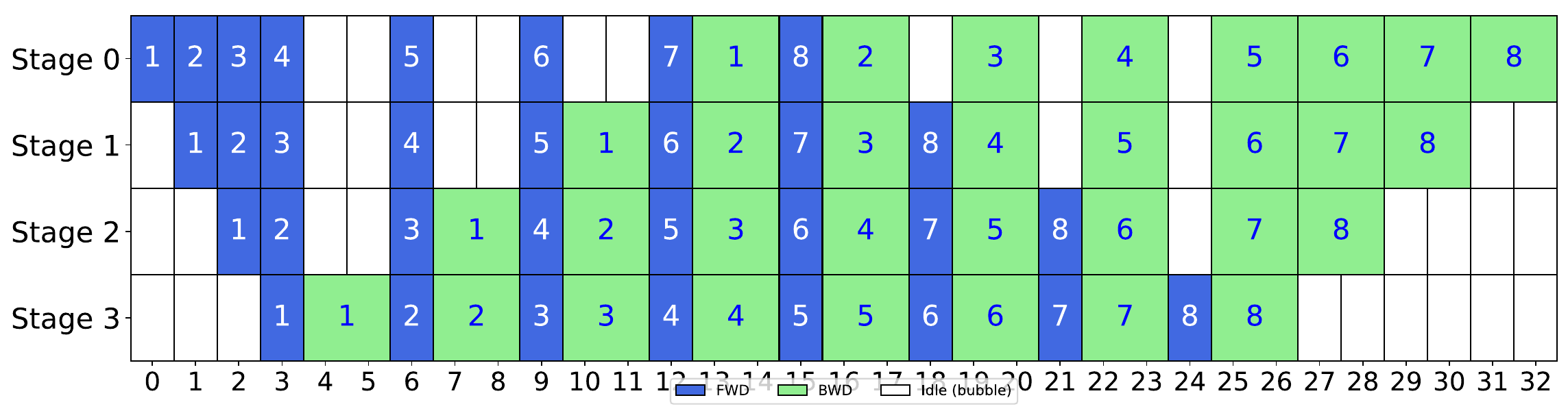}
		\caption{1F1B}
	\end{subfigure}
	\begin{subfigure}{\linewidth}
		\includegraphics[width=\linewidth]{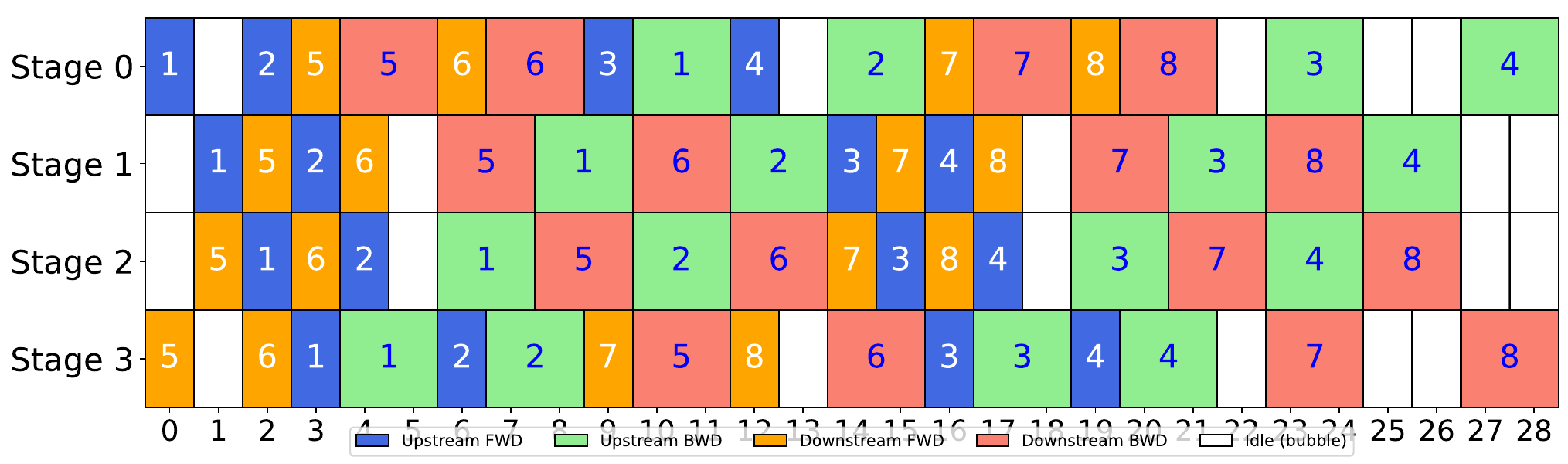}
		\caption{Chimera}
	\end{subfigure}
	\begin{subfigure}{\linewidth}
		\includegraphics[width=\linewidth]{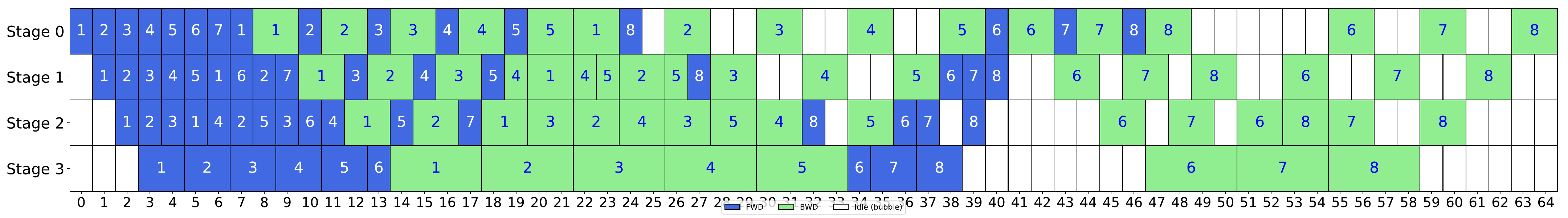}
		\caption{Hanayo}
	\end{subfigure}
	\caption{Illustration of the tabular schedule abstraction for the schedules
	under test with 4 workers operating on 8 microbatches. Rows correspond to
	pipeline workers and columns to discrete schedule slots. Each non-empty cell
	encodes the execution of one microbatch phase on one worker, while empty
	cells represent idle time. Forward compute phases are blue (and orange for
	Chimeras anti-cyclical pipeline) and backward compute phases are colored in
	green (and red, again, for Chimera). The representation makes structural
	properties such as fill, drain, and phase ordering explicit before
	hardware-dependent timing is introduced.} \label{fig:schedule_abstraction}
\end{figure}

Rows in the table correspond to \emph{workers}, not necessarily to individual
accelerators. A worker is treated as an abstract execution unit that may
internally contain additional forms of parallelism, such as tensor or expert
parallelism. The schedule abstraction therefore captures pipeline-level
coordination without exposing worker-internal implementation details. A table
is considered valid if each worker-time slot contains at most one phase, the
causal ordering of phases for every microbatch is preserved, and all required
phases are eventually scheduled. Overall, the tabular abstraction serves as a
compact intermediate representation between abstract schedule design and
communication-aware execution modeling.
Example schedule tables are shown in Figure~\ref{fig:schedule_abstraction}.

\subsection{From schedule tables to execution graphs}
\label{sec:tables_to_graphs}

The tabular schedule abstraction defines what is executed on each worker
and in which discrete slot, but it does not yet encode the explicit dependency
structure required for communication-aware simulation. To obtain such a
representation, each schedule table is translated into an execution graph whose
nodes represent compute and communication events and whose edges capture the
causal dependencies between them.

The translation proceeds in two steps. First, the table is traversed row-wise
to derive the \emph{worker-local execution order}. Each non-empty cell is mapped
to the corresponding compute block for the scheduled microbatch phase, and
consecutive cells on the same worker induce precedence edges in the execution
graph. In this way, the table determines the local sequencing of forward,
backward, optimizer, and optional recomputation phases without requiring
schedule-specific construction logic.

Second, \emph{cross-worker dependencies} are extracted from the phase semantics.
Whenever adjacent pipeline workers process the same microbatch, the upstream
forward phase must produce activations before the downstream forward phase can
begin, and the downstream backward phase must produce gradients before the
upstream backward phase can continue. These dependencies are realized by
inserting explicit send/receive communication nodes between the corresponding
compute blocks. As a result, the execution graph captures not only worker-local
phase order, but also the data movement implied by the schedule.

This construction makes the schedule table the single structural source of
truth for simulation. Different schedule families share the same graph
construction pipeline; they differ only in how the table is populated and, if
needed, in a small number of schedule-specific assumptions such as parameter
duplication or placement rules. The resulting execution graph can then be
annotated with compute and communication costs under a chosen system model and
used to derive runtime, overlap, and memory-related metrics.

\begin{figure}
	\centering
	\includegraphics[width=\linewidth]{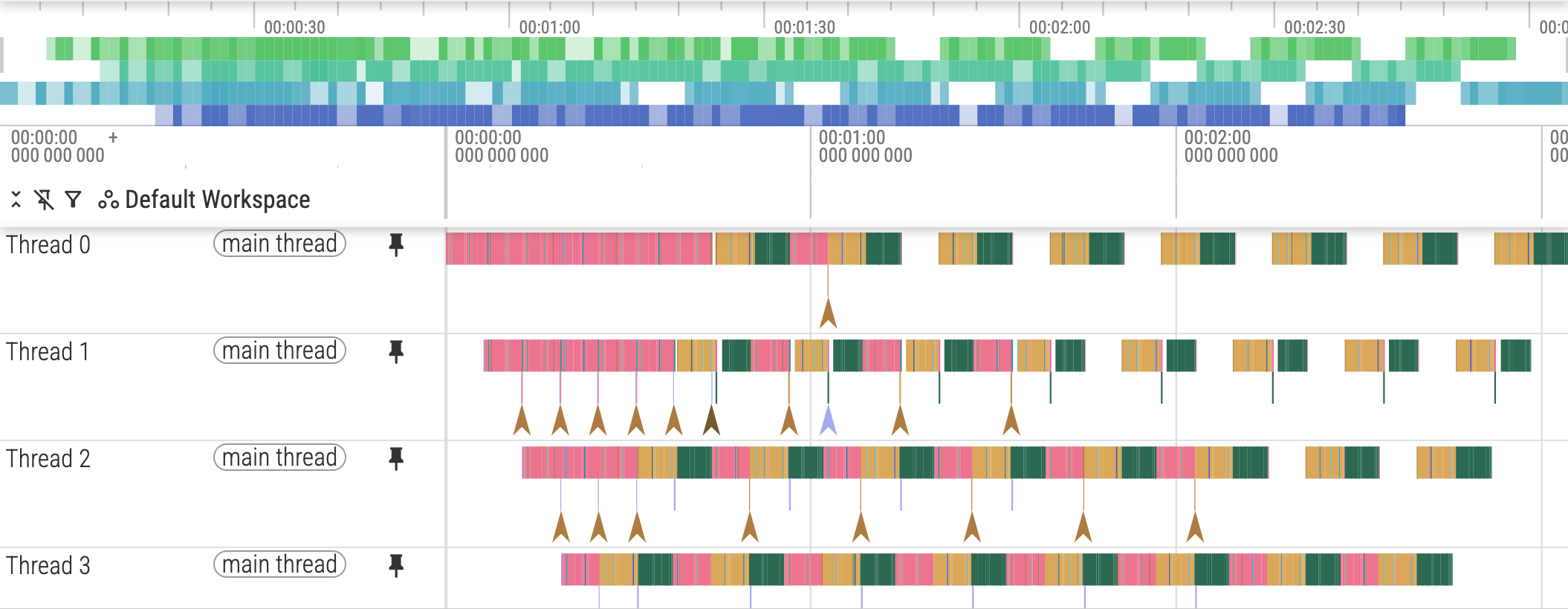}
	\includegraphics[width=\linewidth]{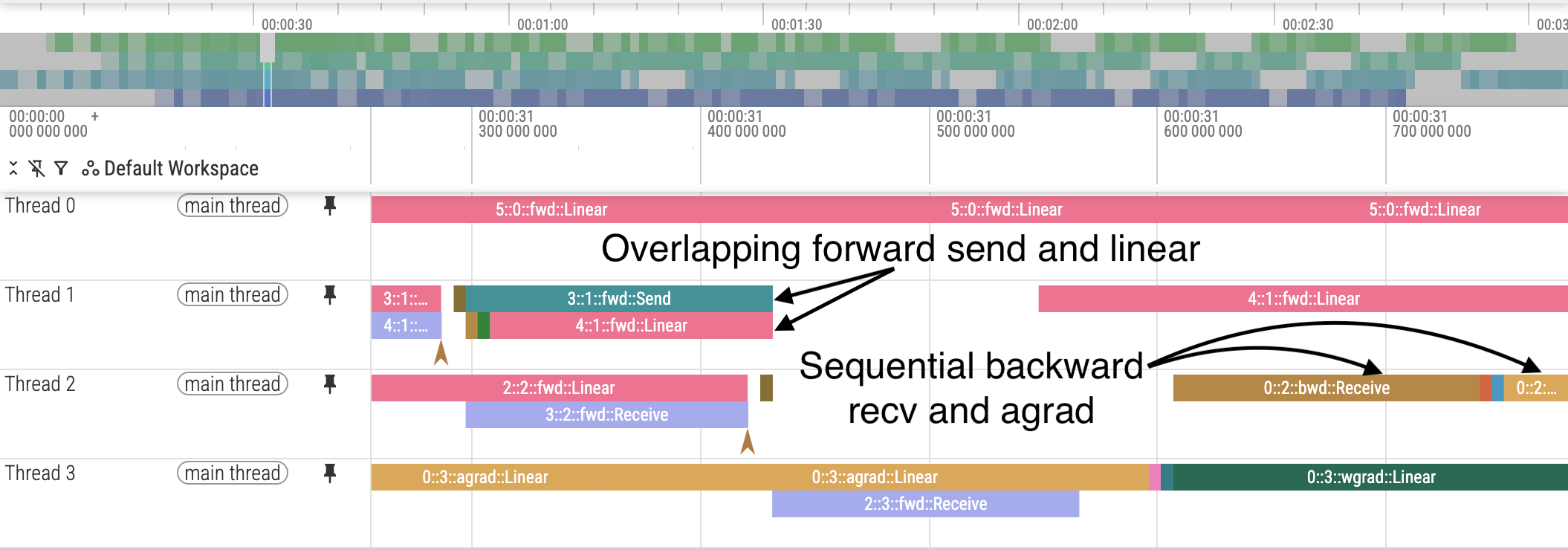}
\caption{Timeline trace generated from the execution graph for the 1F1B
	schedule in Figure~\ref{fig:schedule_abstraction}. The zoomed view (bottom) 
	illustrates overlap between independent forward communication and computation, 
	as well as a dependency-limited backward receive before the agrad phase.}
	\label{fig:timeline}
\end{figure}

\subsection{Three evaluation levels}
\label{sec:three_evaluation_levels}

A central premise of this work is that pipeline schedules should not be
evaluated at only a single level of abstraction. Different abstractions expose
different properties of a schedule, and conclusions that appear stable at one
level need not remain valid once additional execution effects are made
explicit. We therefore compare schedules across three complementary evaluation
levels: \emph{formulas}, \emph{idealized schedule tables}, and
\emph{communication-aware execution simulation}.

At the coarsest level, \emph{formula-based reasoning} provides compact
analytical estimates for structural quantities such as bubble ratios, nominal
utilization, or peak activation counts. This level is useful because it offers
immediate intuition about the asymptotic behavior of a schedule and allows very
fast comparison across many configurations. However, it abstracts away
important effects including communication cost, overlap, and dependency-induced
serialization.

The second level operates directly on the \emph{tabular schedule
representation}. In contrast to closed-form expressions, schedule tables make
the concrete microbatch flow explicit and therefore expose worker-local phase
order, fill and drain behavior, and explicit idle slots. They also support
structural extraction of activation-retention intervals and other schedule
properties that are difficult to recover from formulas alone. At the same time,
this level remains intentionally hardware-agnostic: all slots are interpreted
structurally rather than as physical time.

The third and most detailed level is \emph{communication-aware execution
simulation}. Here, the schedule table is translated into an execution graph and
annotated under a chosen compute and communication model. This makes it
possible to capture system-dependent effects such as exposed communication,
limited overlap between compute and communication, and the impact of different
system regimes on runtime and memory behavior. While this level remains a model
rather than a hardware-calibrated predictor, it provides a substantially more
realistic basis for comparing schedules than structural reasoning alone.

Graphculon is a graph based analytical model based on Calculon~\cite{calculon} that allows us to implement the third level of evaluation.
Graphculon uses the annotated compute graph to generate a timeline, as shown in Figure~\ref{fig:timeline}, using a
capacity-based system model. Since the byte volume \(V_{\mathrm{net}}\) of the communication
is known from the graph, communication time \(t_{\mathrm{comm}}\) can be
calculated following the Hockney model:
\begin{equation}
	t_{\mathrm{comm}} = \frac{V_{\mathrm{net}}}{\mathrm{BW}_{\mathrm{net}}} + L_{\mathrm{net}} ~,
	\label{eq:communication_time}
\end{equation}
using the bandwidth \(\mathrm{BW}_{\mathrm{net}}\) and the latency \(L_{\mathrm{net}}\).

The compute time \(t_{\mathrm{comp}}\) is obtained using a roofline based
model
\begin{equation}
	t_{\mathrm{comp}} = \max\left(\frac{F}{\mathrm{TP} \cdot e_{\mathrm{c}}} + L_{\mathrm{c}},\; \frac{V_{\mathrm{m}}}{\mathrm{BW}_{\mathrm{m}} \cdot e_{\mathrm{m}}} + L_{\mathrm{m}}\right) ~,
	\label{eq:compute_time}
\end{equation}
where a memory-bound workload will be approximated based on the memory time
defined by the byte volume \(V_{\mathrm{m}}\), memory bandwidth \(\mathrm{BW}_\mathrm{m}\) and latency \(L_{\mathrm{m}}\)
and for a compute-bound workload we swap bandwidth for peak throughput
\(\mathrm{TP}\) and the byte volume for the number of FLOPs \(F\) and add a
startup latency of \(L_{\mathrm{c}}\). Two empirical efficiency terms model
additional hardware effects. \(e_{\mathrm{c}}\) estimates the percentage of
achieved compute performance of the peak throughput based on the number of
FLOPs, as does \(e_{\mathrm{m}}\) for the memory based on the byte volume.

Using these three levels side by side serves two purposes. First, it allows us
to separate \emph{structure-driven} conclusions from \emph{system-driven}
conclusions. Second, it enables us to test whether schedule rankings are
stable across abstraction levels or whether they change once communication and
dependencies are modeled explicitly. This multi-abstraction perspective forms
the methodological basis of our evaluation.

\subsection{Metrics}
\label{sec:metrics}

Our evaluation focuses on two classes of metrics: \emph{utilization- and
runtime-related metrics} and \emph{memory-related metrics}. Together, these
capture the main trade-off space of synchronous pipeline schedules.

The first class characterizes how effectively a schedule keeps pipeline workers
busy. At the formula and schedule-table levels, this is expressed through
structural quantities such as \emph{bubble ratio}, \emph{worker utilization},
and \emph{schedule length in slots}. These metrics capture the amount of idle
time induced by pipeline fill, steady-state execution, and drain, and therefore
provide a hardware-agnostic view of how efficiently a schedule uses the
pipeline. At the simulation level, the corresponding metric is the
\emph{modeled execution time} of one training step, which incorporates explicit
compute, communication, and dependency effects under a chosen system model.

The second class captures \emph{per-device memory pressure}. Here, the most
important quantity is the \emph{peak memory requirement} on the bottleneck
worker. We distinguish in particular between \emph{activation memory}, which is
strongly schedule-dependent through activation-retention intervals, and
persistent memory terms such as parameters, gradients, and optimizer state.
This distinction is important because some schedules, such as bidirectional
variants with parameter duplication, may reduce activation pressure while
increasing persistent memory requirements. Accordingly, we consider both
schedule-specific activation peaks and the resulting overall per-device memory
footprint.

These metrics are compared across abstraction levels. 
Bubble ratios and slot-based schedule lengths capture structural properties of a schedule, whereas simulated execution time captures system-dependent behavior. 
Likewise, activation-retention patterns are visible already at the table level, but their practical impact must be
interpreted together with the full memory composition of the worker. 
Our goal is therefore not to collapse all levels into a single scalar score, but to analyze which conclusions remain stable across levels and which emerge only once communication-aware execution is modeled explicitly.
 \section{Experimental Setup}\label{sec:experimental_setup}

In our studies we analyze a Megatron\cite{shoeybi2019megatron} style model with
128 transformer blocks, a hidden dimension \(d_{\mathrm{hidden}} = 4096\), a
number of attention heads \(n_{\mathrm{heads}} = 80\), sequence length
\(s=4096\), and GELU\cite{hendrycks2016gelu} nonlinearity. We further assume a
fixed global minibatch size. Changes to the number of microbatches \(B\)
therefore change the computational work per microbatch. This study focuses on
pipeline parallelism exclusively. The number of stages is denoted \(S\). We do
not vary data, tensor, or expert parallelism and set these domains to 1.

\subsection{Schedules under study} 

In our experiments we evaluate GPipe, 1F1B, Chimera, and Hanayo, as described
in Section~\ref{sec:schedules}. The Hanayo schedule is specifically designed
for \(S = B\) configuration and therefore excluded from the more exhaustive
microbatch sweeps performed for the other schedules.

\subsection{Modeled systems and regimes}

To demonstrate system effects we model a 3 by 3 grid of configurations. This
baseline system configuration models an NVIDIA DGX H100 compute node with
roughly \SI{1}{\peta FLOPs} of compute, \SI{34}{\tera\byte\per\second} total
memory bandwidth with a memory latency of \SI{50}{\nano\second}, and a
\SI{50}{\giga\byte\per\second} Infiniband interconnect with a latency of
\SI{500}{\nano\second}. We then scale both the compute and network components
by a factor of 10 in both directions to obtain a fast and slow compute system,
\texttt{fast\_cp} and \texttt{slow\_cp} respectively, and fast and slow network
system, \texttt{fast\_nw} and \texttt{slow\_nw}. Both throughput/bandwidth and
latency are scaled.

Although this 10$\times$ scaling is stylized, it is not intended to be purely
hypothetical: modern rack-scale scale-up fabrics such as NVIDIA NVL72 expose
aggregate communication bandwidth far beyond conventional inter-node
InfiniBand. We therefore interpret the fast-network regime as a
communication-favorable envelope rather than as a direct model of a specific
link technology.

\subsection{Scope and assumptions}

Our study is of a comparative nature. Rather than predicting exact numbers for
an explicit hardware target we want to explore the effects hardware has on
communication schedules otherwise hidden at higher levels of abstraction.
Additionally, we limit the scope to synchronous schedules with identical
training semantics. Otherwise, comparisons would not be meaningful
without also reporting convergence and accuracy metrics, requiring an actual
training run which is out of the scope of this work.
 \section{Comparing Pipeline Schedules Across Abstraction Levels }

This section compares the schedules listed in
Section~\ref{sec:experimental_setup} under the levels of abstraction listed in
Section~\ref{sec:method}. We first compare the analytical formulae to the time
slot table view and move to communication-aware simulation afterward.

\subsection{Structural comparison without simulation}

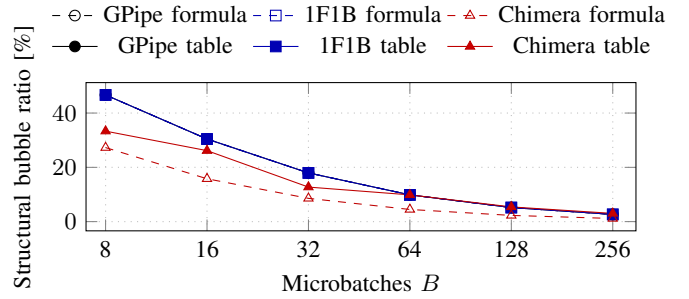
\begin{figure}
	\centering
	\begin{tikzpicture}
\begin{axis}[
	width=\columnwidth,
	height=0.4\columnwidth,
	xmode=log,
	log basis x=2,
	xmin=7,
	xmax=300,
	xtick={8,16,32,64,128,256},
	xticklabels={8,16,32,64,128,256},
	xlabel={Microbatches $B$},
	ylabel={Structural bubble ratio [\%]},
	ymajorgrids=true,
	xmajorgrids=true,
	grid style={dotted},
	tick label style={font=\small},
	label style={font=\small},
	legend style={
		at={(0.5,1.1)},
		anchor=south,
		legend columns=2,
		transpose legend,
		draw=none,
		fill=none,
		font=\small
	},
	clip=false,
]

\addplot+[black, dashed, mark=o, mark options={solid, fill=white, draw=black}]
	table[col sep=comma, x=B, y=gpipe_formula]
	{data/formula_comparison.dat};
\addlegendentry{GPipe formula}

	\addplot+[black, solid, mark=*, mark options={solid, fill=black, draw=black}]
	table[col sep=comma, x=B, y=gpipe_table]
	{data/formula_comparison.dat};
\addlegendentry{GPipe table}

\addplot+[blue!70!black, dashed, mark=square, mark options={solid, fill=white, draw=blue!70!black}]
	table[col sep=comma, x=B, y=onef1b_formula]
	{data/formula_comparison.dat};
\addlegendentry{1F1B formula}

\addplot+[blue!70!black, solid, mark=square*]
	table[col sep=comma, x=B, y=onef1b_table]
	{data/formula_comparison.dat};
\addlegendentry{1F1B table}

\addplot+[red!75!black, dashed, mark=triangle, mark options={solid, fill=white, draw=red!75!black}]
	table[col sep=comma, x=B, y=chimera_formula]
	{data/formula_comparison.dat};
\addlegendentry{Chimera formula}

\addplot+[red!75!black, solid, mark=triangle*]
	table[col sep=comma, x=B, y=chimera_table]
	{data/formula_comparison.dat};
\addlegendentry{Chimera table}

\end{axis}
\end{tikzpicture}
 	\caption{Structural bubble ratio comparison between analytical formulae and
	time slot table abstraction for GPipe, 1F1B, and Chimera at number of
	pipeline stages \(S = 8\). As expected, GPipe and 1F1B are equivalent in bubble ratio.
	This also translates directly to the tabular abstraction. For Chimera we
	observe the analytical formula to be more optimistic than the instantiated
	table. 
}\label{fig:formula_comparison}
\end{figure}

Comparing the closed-form formulae to the instantiated time slot table, as
shown in Figure~\ref{fig:formula_comparison}, already reveals some
discrepancies for the Chimera schedule. GPipe and 1F1B not only behave the same
in the analytical model, which is expected of course, but also match the
tabular representation perfectly. Note that they overlap perfectly in the figure.
For Chimera, however, we observe slightly
larger bubble ratios using the tabular abstraction than we obtain from the
formula. The effect is more pronounced for a smaller number of microbatches but
still holds true for \(B = 256\), albeit at a significantly smaller difference.
We observe the same effect when varying the number of stages. At \((S,B) =
(8,16)\) we obtain bubble ratios of 16\% vs. 26\% from formula and table
respectively. At \((S, B) = (4, 16)\) we similarly observe values of 6\% vs.
13\%.

Because we use the same timing assumptions (\(t_{\mathrm{bwd}} =
2t_{\mathrm{fwd}}\)) in both cases and preserve the same bidirectional
execution pattern from the original paper we conclude that the two abstraction levels disagree for Chimera. However, in all cases Chimera
outperforms GPipe and 1F1B with a lower bubble ratio.

\subsection{Runtime under communication-aware modeling}
\label{sec:schedule_comparison}

When we introduce explicit communication through simulation via Graphculon we
immediately observe how side effects can alter the results. While Chimera
consistently shows a lower or at least equal bubble ratio using the tabular
abstraction, we observe, especially for network-bound systems, that GPipe and
1F1B offer a better idle to compute time ratio and overall runtime at a higher
microbatch count. Simulating systems from network-bound to compute-bound
confirms our thesis that the quality of schedule highly depends on the system
configuration and training hyperparameters
used, as we will elaborate in the following.

\begin{figure*}
	\centering
	{
\begin{tabular}{cc}
  \tikz[baseline=-0.6ex]{
    \draw[blue!70!black, line width=0.8pt] (0,0) -- (0.42,0);
    \fill[blue!70!black] (0.21,0) ++(-1.8pt,-1.8pt) rectangle ++(3.6pt,3.6pt);
  }\ GPipe/1F1B
  &
  \tikz[baseline=-0.6ex]{
    \draw[red!75!black, line width=0.8pt] (0,0) -- (0.42,0);
    \fill[red!75!black] (0.21,2.1pt) -- ++(-2.3pt,-4.0pt) -- ++(4.6pt,0) -- cycle;
  }\ Chimera
\end{tabular}
}

\vspace{0.35em}
\begin{tikzpicture}
\begin{groupplot}[
  group style={group size=3 by 2, horizontal sep=0.9cm, vertical sep=0.6cm},
  width=0.35\textwidth,
  height=0.15\textwidth,
  xmode=log,
  log basis x=2,
  xmin=7,
  xmax=300,
  xtick={8,16,32,64,128,256},
  xticklabels={8,16,32,64,128,256},
  xmajorgrids=true,
  ymajorgrids=true,
  grid style={dotted},
  title style={align=center},
  clip=false,
]

\nextgroupplot[
  title={Slow network fast compute},
  ylabel={$\beta_{\mathrm{idle}}$ [\%]}
]
\addplot+[black, dashed, mark=none]
  table[col sep=comma, x=B, y=slow_gpipe]
  {data/timeline_comparison_bubble.dat};
\addplot+[blue!70!black, solid, mark=none]
  table[col sep=comma, x=B, y=slow_onef1b]
  {data/timeline_comparison_bubble.dat};
\addplot+[red!75!black, solid, mark=none]
  table[col sep=comma, x=B, y=slow_chimera]
  {data/timeline_comparison_bubble.dat};
\addplot+[only marks, black, mark=o, mark options={solid, fill=white, draw=black}]
  table[col sep=comma, x=B, y=slow_gpipe]
  {data/timeline_comparison_bubble.dat};
\addplot+[only marks, blue!70!black, mark=square*, mark options={solid, fill=blue!70!black, draw=blue!70!black}]
  table[col sep=comma, x=B, y=slow_onef1b]
  {data/timeline_comparison_bubble.dat};
\addplot+[only marks, red!75!black, mark=triangle*, mark options={solid, fill=red!75!black, draw=red!75!black}]
  table[col sep=comma, x=B, y=slow_chimera]
  {data/timeline_comparison_bubble.dat};

\nextgroupplot[
  title={Baseline system},
  ylabel={}
]
\addplot+[black, dashed, mark=none]
  table[col sep=comma, x=B, y=mid_gpipe]
  {data/timeline_comparison_bubble.dat};
\addplot+[blue!70!black, solid, mark=none]
  table[col sep=comma, x=B, y=mid_onef1b]
  {data/timeline_comparison_bubble.dat};
\addplot+[red!75!black, solid, mark=none]
  table[col sep=comma, x=B, y=mid_chimera]
  {data/timeline_comparison_bubble.dat};
\addplot+[only marks, black, mark=o, mark options={solid, fill=white, draw=black}]
  table[col sep=comma, x=B, y=mid_gpipe]
  {data/timeline_comparison_bubble.dat};
\addplot+[only marks, blue!70!black, mark=square*, mark options={solid, fill=blue!70!black, draw=blue!70!black}]
  table[col sep=comma, x=B, y=mid_onef1b]
  {data/timeline_comparison_bubble.dat};
\addplot+[only marks, red!75!black, mark=triangle*, mark options={solid, fill=red!75!black, draw=red!75!black}]
  table[col sep=comma, x=B, y=mid_chimera]
  {data/timeline_comparison_bubble.dat};

\nextgroupplot[
  title={Fast network slow compute},
  ylabel={}
]
\addplot+[black, dashed, mark=none]
  table[col sep=comma, x=B, y=fast_gpipe]
  {data/timeline_comparison_bubble.dat};
\addplot+[blue!70!black, solid, mark=none]
  table[col sep=comma, x=B, y=fast_onef1b]
  {data/timeline_comparison_bubble.dat};
\addplot+[red!75!black, solid, mark=none]
  table[col sep=comma, x=B, y=fast_chimera]
  {data/timeline_comparison_bubble.dat};
\addplot+[only marks, black, mark=o, mark options={solid, fill=white, draw=black}]
  table[col sep=comma, x=B, y=fast_gpipe]
  {data/timeline_comparison_bubble.dat};
\addplot+[only marks, blue!70!black, mark=square*, mark options={solid, fill=blue!70!black, draw=blue!70!black}]
  table[col sep=comma, x=B, y=fast_onef1b]
  {data/timeline_comparison_bubble.dat};
\addplot+[only marks, red!75!black, mark=triangle*, mark options={solid, fill=red!75!black, draw=red!75!black}]
  table[col sep=comma, x=B, y=fast_chimera]
  {data/timeline_comparison_bubble.dat};

\nextgroupplot[
  ylabel={$T_{\mathrm{sim}}$ [s]}
]
\addplot+[black, dashed, mark=none]
  table[col sep=comma, x=B, y=slow_gpipe]
  {data/timeline_comparison_runtime.dat};
\addplot+[blue!70!black, solid, mark=none]
  table[col sep=comma, x=B, y=slow_onef1b]
  {data/timeline_comparison_runtime.dat};
\addplot+[red!75!black, solid, mark=none]
  table[col sep=comma, x=B, y=slow_chimera]
  {data/timeline_comparison_runtime.dat};
\addplot+[only marks, black, mark=o, mark options={solid, fill=white, draw=black}]
  table[col sep=comma, x=B, y=slow_gpipe]
  {data/timeline_comparison_runtime.dat};
\addplot+[only marks, blue!70!black, mark=square*, mark options={solid, fill=blue!70!black, draw=blue!70!black}]
  table[col sep=comma, x=B, y=slow_onef1b]
  {data/timeline_comparison_runtime.dat};
\addplot+[only marks, red!75!black, mark=triangle*, mark options={solid, fill=red!75!black, draw=red!75!black}]
  table[col sep=comma, x=B, y=slow_chimera]
  {data/timeline_comparison_runtime.dat};

\nextgroupplot[
  ylabel={}
]
\addplot+[black, dashed, mark=none]
  table[col sep=comma, x=B, y=mid_gpipe]
  {data/timeline_comparison_runtime.dat};
\addplot+[blue!70!black, solid, mark=none]
  table[col sep=comma, x=B, y=mid_onef1b]
  {data/timeline_comparison_runtime.dat};
\addplot+[red!75!black, solid, mark=none]
  table[col sep=comma, x=B, y=mid_chimera]
  {data/timeline_comparison_runtime.dat};
\addplot+[only marks, black, mark=o, mark options={solid, fill=white, draw=black}]
  table[col sep=comma, x=B, y=mid_gpipe]
  {data/timeline_comparison_runtime.dat};
\addplot+[only marks, blue!70!black, mark=square*, mark options={solid, fill=blue!70!black, draw=blue!70!black}]
  table[col sep=comma, x=B, y=mid_onef1b]
  {data/timeline_comparison_runtime.dat};
\addplot+[only marks, red!75!black, mark=triangle*, mark options={solid, fill=red!75!black, draw=red!75!black}]
  table[col sep=comma, x=B, y=mid_chimera]
  {data/timeline_comparison_runtime.dat};

\nextgroupplot[
  ylabel={}
]
\addplot+[black, dashed, mark=none]
  table[col sep=comma, x=B, y=fast_gpipe]
  {data/timeline_comparison_runtime.dat};
\addplot+[blue!70!black, solid, mark=none]
  table[col sep=comma, x=B, y=fast_onef1b]
  {data/timeline_comparison_runtime.dat};
\addplot+[red!75!black, solid, mark=none]
  table[col sep=comma, x=B, y=fast_chimera]
  {data/timeline_comparison_runtime.dat};
\addplot+[only marks, black, mark=o, mark options={solid, fill=white, draw=black}]
  table[col sep=comma, x=B, y=fast_gpipe]
  {data/timeline_comparison_runtime.dat};
\addplot+[only marks, blue!70!black, mark=square*, mark options={solid, fill=blue!70!black, draw=blue!70!black}]
  table[col sep=comma, x=B, y=fast_onef1b]
  {data/timeline_comparison_runtime.dat};
\addplot+[only marks, red!75!black, mark=triangle*, mark options={solid, fill=red!75!black, draw=red!75!black}]
  table[col sep=comma, x=B, y=fast_chimera]
  {data/timeline_comparison_runtime.dat};

\end{groupplot}
\node at ($(group c1r2.south)!0.5!(group c3r2.south)+(0,-0.9cm)$) {Microbatches $B$};
\end{tikzpicture}
  \caption{Simulated runtime
	\(T_{\mathrm{sim}}\) and idle time ratio \(\beta_{\mathrm{idle}}\) comparison
	of GPipe, 1F1B, and Chimera processing a single minibatch on \(S=8\) pipeline
	stages for varying number of microbatches on 3 different systems: one network
	bound, one balanced, and one compute-bound.
}\label{fig:timeline_comparison}
\end{figure*}
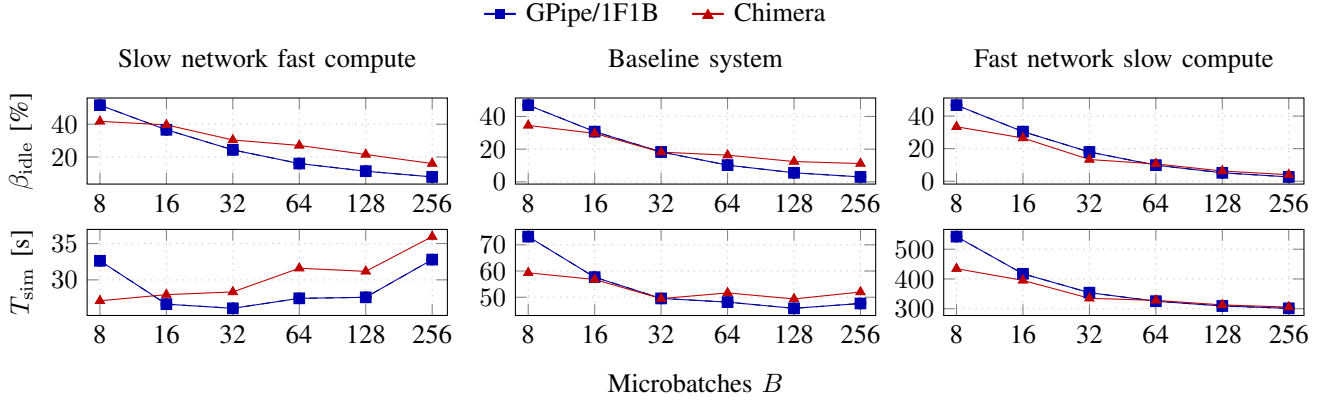

Figure~\ref{fig:timeline_comparison} shows both simulated runtime and
idle time for processing a minibatch and over 8 pipeline stages and various
numbers of microbatches on three different systems.
On network-bound systems we see that increasing the number of microbatches will
even lead to longer runtimes. This intuitively makes sense as this leads to
more communication overall. The more communication-heavy Chimera
schedule is more affected by this than GPipe and 1F1B. On the baseline system
the difference is less pronounced, but beyond 32 microbatches GPipe and 1F1B
show better idle and runtime on a system. Finally, on a system with a very
fast network compared to compute Chimera outperforms GPipe and 1F1B until a
microbatch count of 64. Beyond that, the schedules converge.

\subsection{Memory behavior across schedules}

One important factor not mentioned so far is the per-node peak memory required
for the schedule in question. Even with recent development in datacenter-grade
GPUs they are notoriously short on memory. In most cases peak memory
consumption will dictate whether a schedule is feasible or not. We have already
observed the structural and runtime differences among the schedules. They
behave equally different in the memory domain.

\begin{figure}
	\centering
	{
\begin{tabular}{ccc}
  \tikz[baseline=-0.6ex]{
    \draw[black, line width=0.6pt] (0,0) -- (0.42,0);
    \filldraw[fill=black, draw=black, line width=0.8pt] (0.21,0) circle[radius=1.8pt];
  }\ GPipe
  &
  \tikz[baseline=-0.6ex]{
    \draw[blue!70!black, line width=0.8pt] (0,0) -- (0.42,0);
    \fill[blue!70!black] (0.21,0) ++(-1.8pt,-1.8pt) rectangle ++(3.6pt,3.6pt);
  }\ 1F1B
  &
  \tikz[baseline=-0.6ex]{
    \draw[red!75!black, line width=0.8pt] (0,0) -- (0.42,0);
    \fill[red!75!black] (0.21,2.1pt) -- ++(-2.3pt,-4.0pt) -- ++(4.6pt,0) -- cycle;
  }\ Chimera
\end{tabular}
}

\vspace{0.35em}
\begin{tikzpicture}
\begin{groupplot}[
  group style={group size=2 by 1, horizontal sep=1.2cm},
  width=0.5\columnwidth,
  height=0.35\columnwidth,
  xmode=log,
  ymode=log,
  log basis x=2,
  xmin=7,
  xmax=300,
  ymin=0.3,
  ymax=32,
  xtick={8,16,32,64,128,256},
  xticklabels={8,16,32,64,128,256},
  xmajorgrids=true,
  ymajorgrids=true,
  grid style={dotted},
  tick label style={font=\small},
  label style={font=\small},
  title style={font=\small},
  clip=false,
]

\nextgroupplot[title={$S=4$}]
\addplot+[black, mark=none]
  table[col sep=comma, x=B, y=gpipe_s4]
  {data/memory.dat};
\addplot+[blue!70!black, solid, mark=none]
  table[col sep=comma, x=B, y=onef1b_s4]
  {data/memory.dat};
\addplot+[red!75!black, solid, mark=none]
  table[col sep=comma, x=B, y=chimera_s4]
  {data/memory.dat};
\addplot+[only marks, black, mark=o, mark options={solid, fill=black, draw=black}]
  table[col sep=comma, x=B, y=gpipe_s4]
  {data/memory.dat};
\addplot+[only marks, blue!70!black, mark=square*, mark options={solid, fill=blue!70!black, draw=blue!70!black}]
  table[col sep=comma, x=B, y=onef1b_s4]
  {data/memory.dat};
\addplot+[only marks, red!75!black, mark=triangle*, mark options={solid, fill=red!75!black, draw=red!75!black}]
  table[col sep=comma, x=B, y=chimera_s4]
  {data/memory.dat};

\nextgroupplot[title={$S=8$}, ylabel={}]
\addplot+[black, mark=none]
  table[col sep=comma, x=B, y=gpipe_s8]
  {data/memory.dat};
\addplot+[blue!70!black, solid, mark=none]
  table[col sep=comma, x=B, y=onef1b_s8]
  {data/memory.dat};
\addplot+[red!75!black, solid, mark=none]
  table[col sep=comma, x=B, y=chimera_s8]
  {data/memory.dat};
\addplot+[only marks, black, mark=o, mark options={solid, fill=black, draw=black}]
  table[col sep=comma, x=B, y=gpipe_s8]
  {data/memory.dat};
\addplot+[only marks, blue!70!black, mark=square*, mark options={solid, fill=blue!70!black, draw=blue!70!black}]
  table[col sep=comma, x=B, y=onef1b_s8]
  {data/memory.dat};
\addplot+[only marks, red!75!black, mark=triangle*, mark options={solid, fill=red!75!black, draw=red!75!black}]
  table[col sep=comma, x=B, y=chimera_s8]
  {data/memory.dat};

\end{groupplot}
\node[rotate=90, anchor=center] at ($(group c1r1.west)+(-0.85cm,0)$) {Activation mem [TiB]};
\node at ($(group c1r1.south west)!0.5!(group c2r1.south east)+(0,-0.85cm)$) {Microbatches $B$};
\end{tikzpicture}
 	\caption{Peak per-device activation memory consumption for 4 and 8 pipeline
	stages. Due to its structure GPipe does not change with varying microbatches,
	since after all forward passes are complete there always is a point in time
	where the full minibatch of activations resides in memory.}\label{fig:memory}
\end{figure}
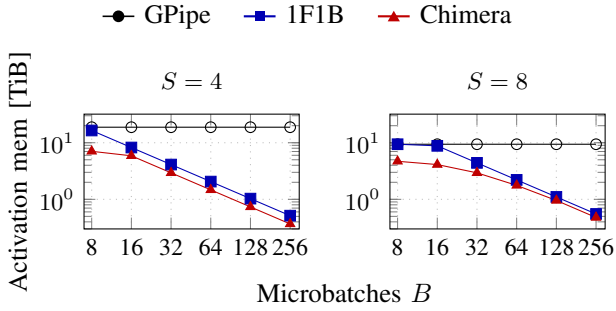

Figure~\ref{fig:memory} shows the peak per-device memory for different
microbatch counts and 4 and 8 pipeline stages. GPipe's memory consumption is
invariant to the microbatch count. Due to its nature to eagerly compute all
forward passes, there exists a point in time, after the last forward pass, where
the whole minibatch of activations accumulates until they can be deallocated
after the respective backward phase.
1F1B reduces the activation-memory peak relative to GPipe by shortening
activation-retention intervals once steady state is reached. Chimera lowers
activation pressure further in the evaluated settings, but this advantage must
be interpreted together with its higher communication demand and possible
persistent-memory overheads due to parameter duplication.

\subsection{Hanayo in its restricted regime}
\label{sec:hanayo_restricted}

We additionally compare Chimera against two-wave Hanayo in the restricted
$(S,B)=(8,8)$ regime, which corresponds to the intended operating point of the
schedule. The comparison is summarized in
Table~\ref{tab:hanayo_comparison}. In this setting, Hanayo outperforms Chimera
in eight of the nine modeled systems. Across fast- and medium-network regimes,
the runtime improvement is highly consistent, typically in the range of
approximately $11\%$--$14\%$, and is accompanied by noticeably lower idle-time
ratios. For example, on the baseline system, runtime decreases from
$59.32\,\mathrm{s}$ to $51.79\,\mathrm{s}$, corresponding to a $12.69\%$
improvement, while idle time drops from $34.51\%$ to $24.99\%$.

At the same time, this advantage is not robust across all communication
regimes. As network performance degrades, the benefit of Hanayo becomes
smaller and may disappear entirely. In the \texttt{slow\_nw\_mid\_cp} regime,
Hanayo improves runtime by only $2.33\%$, while in the most
communication-constrained \texttt{slow\_nw\_fast\_cp} regime it becomes
$12.32\%$ slower than Chimera and exhibits a higher idle ratio. We therefore
interpret Hanayo as effective in its intended restricted operating point, but
still sensitive to communication bottlenecks. This result is consistent with
the broader conclusion of the paper: even schedules with favorable structural
or restricted-regime behavior do not admit a universal ranking independent of
the system model.

\begin{table}[htbp]
\centering
\caption{Comparison of Chimera (C) and two-wave Hanayo (H) in the restricted
$(S,B)=(8,8)$ regime. Negative $\Delta T$ means that Hanayo is faster than
Chimera.}
\label{tab:hanayo_comparison}
\begin{tabular}{lrrrrr}
\hline
& \multicolumn{2}{c}{$\beta_{\mathrm{idle}}$ [\%]} &
  \multicolumn{2}{c}{$T_{\mathrm{sim}}$ [s]} &
  \\
\cline{2-3} 
\cline{4-5}
System & \multicolumn{1}{c}{C} & \multicolumn{1}{c}{H} & \multicolumn{1}{c}{C} & \multicolumn{1}{c}{H} & \multicolumn{1}{c}{$\Delta T$ [\%]} \\
\hline
\texttt{fast\_nw\_fast\_cp} & 35.67 & 25.47 & 24.56  & 21.20  & -13.69 \\
\texttt{fast\_nw\_mid\_cp}  & 34.26 & 23.76 & 59.08  & 50.95  & -13.77 \\
\texttt{fast\_nw\_slow\_cp} & 33.44 & 22.79 & 434.54 & 374.63 & -13.79 \\
\texttt{mid\_nw\_fast\_cp}  & 36.28 & 28.31 & 24.79  & 22.04  & -11.11 \\
\texttt{baseline}           & 34.51 & 24.99 & 59.32  & 51.79  & -12.69 \\
\texttt{mid\_nw\_slow\_cp}  & 33.47 & 22.96 & 434.77 & 375.47 & -13.64 \\
\texttt{slow\_nw\_fast\_cp} & 41.73 & 48.12 & 27.11  & 30.45  & 12.32 \\
\texttt{slow\_nw\_mid\_cp}  & 36.98 & 35.47 & 61.64  & 60.20  & -2.33 \\
\texttt{slow\_nw\_slow\_cp} & 33.83 & 24.65 & 437.09 & 383.87 & -12.18 \\
\hline
\end{tabular}
\end{table}

\subsection{Verdict: schedule rankings are not abstraction-invariant}

Taken together, the results across Figures~\ref{fig:formula_comparison}, \ref{fig:timeline_comparison}, \ref{fig:memory}, and Table~\ref{tab:hanayo_comparison} show that pipeline schedules cannot be ranked reliably from structural reasoning alone. 
Analytical formulae and instantiated schedule tables may suggest one ordering, but communication-aware simulation can change that ordering once exposed communication, overlap constraints, and system bottlenecks are made explicit.

Under the assumptions considered here, GPipe and 1F1B are runtime-equivalent, but 1F1B achieves a lower memory peak and therefore emerges as the stronger unidirectional baseline. 
Chimera is advantageous primarily at low microbatch counts and in communication-favorable regimes, where its bidirectional structure can reduce idle time without being dominated by additional communication overhead. Hanayo is effective in its intended restricted regime, but its advantage weakens and can reverse under communication bottlenecks.
Overall, no schedule is uniformly superior across abstraction levels and system regimes. The quality of a pipeline schedule is therefore meaningful only in the context of the modeled execution environment.
 \section{Case Study: Asymmetric Chimera Placement }
\label{sec:asymmetric_chimera}

In this section we apply the simulation framework presented in
Section~\ref{sec:method} to the Chimera schedule in a what-if analysis. 
This serves primarily as a demonstration of the applicability of the presented abstractions rather than
as a substantial improvement of the existing schedule.
Section~\ref{sec:schedule_comparison} shows that Chimera, in a low microbatch
regime and with sufficient network speeds, offers better run and idle-time
ratios and notably lower memory utilization than GPipe and 1F1B. However, as is
also the case in 1F1B, the memory pressure is not shared equally across nodes.
Typically, the first node in the pipeline has to retain the largest amount of
activations, thereby becoming the bottleneck in the overall execution. This
begs the question whether this bottleneck can be alleviated through shuffling
of intra-pipeline block placement, keeping the total per-device model parameter
count fixed.

The original Chimera schedule features two mirrored pipelines over the same
device set. In practice, both directions follow the same sequence of stages and
distribute blocks uniformly across workers. In theory, nothing disallows a
non-uniform distribution. In this experiment we move work from the early stages
by assigning fewer blocks to the first half of one of the pipelines and more to
the latter half and vice versa for the second pipeline. This maintains a meta
symmetry in which each worker still experiences the same amount of work. This
split creates some constraints on the model, mainly that the number of blocks
be divisible by the number of stages. We thus use \(N=120\) blocks instead
of \(N=128\).

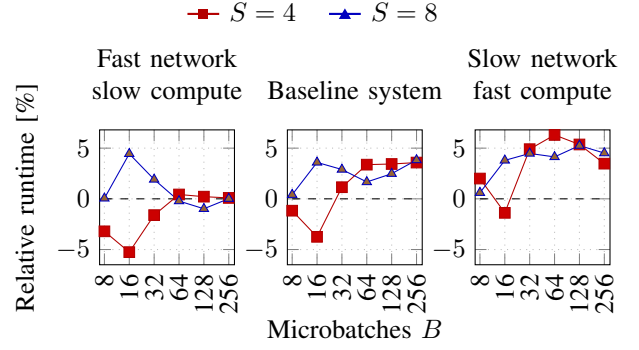
\begin{figure}
	\centering
	{
\begin{tabular}{cc}
  \tikz[baseline=-0.6ex]{
    \draw[red!70!black, line width=0.8pt] (0,0) -- (0.42,0);
    \fill[red!70!black] (0.21,0) ++(-1.8pt,-1.8pt) rectangle ++(3.6pt,3.6pt);
  }\ $S=4$
  &
  \tikz[baseline=-0.6ex]{
    \draw[blue!75!black, line width=0.8pt] (0,0) -- (0.42,0);
    \fill[blue!75!black] (0.21,2.1pt) -- ++(-2.3pt,-4.0pt) -- ++(4.6pt,0) -- cycle;
  }\ $S=8$
\end{tabular}
}

\vspace{0.35em}
\begin{tikzpicture}
\begin{groupplot}[
  group style={group size=3 by 1, horizontal sep=0.7cm},
  width=0.38\columnwidth,
  height=0.38\columnwidth,
  xmode=log,
  log basis x=2,
  xmin=7,
  xmax=300,
  xtick={8,16,32,64,128,256},
  xticklabels={8,16,32,64,128,256},
  ylabel={Relative runtime [\%]},
  ymin=-6.5,
  ymax=6.8,
  ymajorgrids=true,
  xmajorgrids=true,
  grid style={dotted},
  x tick label style={rotate=90},
  label style={},
  title style={ align=center},
  clip=false,
]

\nextgroupplot[title={Fast network\\slow compute}]
  \addplot+[black, dashed, mark=none] coordinates {(8,0) (256,0)};
  \addplot+[red!70!black, solid, mark=square*]
    table[col sep=comma, x=B, y=fast_s4_pct] {data/unbalanced_runtime.dat};

  \addplot+[blue!75!black, solid, mark=triangle*]
    table[col sep=comma, x=B, y=fast_s8_pct] {data/unbalanced_runtime.dat};

\nextgroupplot[title={Baseline system}, ylabel={}]
  \addplot+[black, dashed, mark=none] coordinates {(8,0) (256,0)};
  \addplot+[red!70!black, solid, mark=square*]
    table[col sep=comma, x=B, y=mid_s4_pct] {data/unbalanced_runtime.dat};

  \addplot+[blue!75!black, solid, mark=triangle*]
    table[col sep=comma, x=B, y=mid_s8_pct] {data/unbalanced_runtime.dat};

\nextgroupplot[title={Slow network\\fast compute}, ylabel={}]
  \addplot+[black, dashed, mark=none] coordinates {(8,0) (256,0)};
  \addplot+[red!70!black, solid, mark=square*]
    table[col sep=comma, x=B, y=slow_s4_pct] {data/unbalanced_runtime.dat};

  \addplot+[blue!75!black, solid, mark=triangle*]
    table[col sep=comma, x=B, y=slow_s8_pct] {data/unbalanced_runtime.dat};

\end{groupplot}
\node[] at ($(group c1r1.south)!0.5!(group c3r1.south)+(0,-0.85cm)$) {Microbatches $B$};
\end{tikzpicture}
 	\caption{Relative simulated runtime compared to the baseline Chimera schedule
	on the network-bound, baseline, and
	compute-bound system. Lower is better. Interestingly, for 16 microbatches the
	best and worst results are encountered for the shallow and deeper pipelines
	respectively.}\label{fig:unbalanced_runtime}
\end{figure}

Interestingly we do not observe an improvement in global peak memory, only a
more uniform distribution across workers. However, we unexpectedly do observe
slight improvements in overall runtime. Even more unexpected, there is a strong
correlation with the number of stages. The best improvements can be found in
shallow pipelines in communication-favorable settings, i.e. fast communication
systems. At a depth of \(S=4\) stages we observe a speedup of 5\% compared to
the baseline schedule at lower microbatch counts. Going to more stages (\(S=8\)), we instead observe a slowdown for few microbatches, but a marginal improvement for larger counts. Figure~\ref{fig:unbalanced_runtime} shows the
relative runtime of the unbalanced schedule compared to the baseline balanced
schedule.

Although the observed runtime improvements are modest, this case study
demonstrates the usefulness of simulation-based what-if analysis. The original
goal of reducing peak memory was not achieved, yet the framework still exposed
non-obvious runtime behavior that would not have been apparent from either a
tabular or purely analytical view.
 \section{Conclusion}

We introduced a tabular schedule abstraction and a unified multi-abstraction
methodology for comparing pipeline schedules in distributed LLM training. By
connecting analytical formulae, idealized schedule tables, and
communication-aware execution simulation, the framework separates
structure-driven from system-driven schedule behavior. Our results show that
schedule rankings are not abstraction-invariant: GPipe and 1F1B are
runtime-equivalent under the modeled assumptions, but 1F1B has a lower
activation-memory peak; Chimera is beneficial mainly at low microbatch counts
and in communication-favorable regimes; and Hanayo is effective in its intended
restricted regime but remains sensitive to network bottlenecks. The asymmetric
Chimera case study further shows that plausible schedule modifications may fail
to improve the targeted bottleneck while still revealing non-obvious runtime
effects. Overall, pipeline schedule quality cannot be inferred reliably from
structural metrics alone, but must be evaluated under an explicit system model
that captures communication, dependency, and memory interactions.

For future work, we are particularly interested in extending the framework toward communication-aware energy modeling~\cite{braun2021,zahn2019,tarragamoreno2025} and in calibrating the computational model across a broader range of hardware platforms~\cite{emonds2023cudasap,braun2021}. 
Another important direction is support for more complex schedule classes, such as interleaved and zero-bubble pipelining~\cite{qi2024zero,narayanan2021efficient}. 
Compression techniques, especially for activation state~\cite{barley2023,barley2024}, are also promising, although their evaluation will require modeling quality-efficiency trade-offs.
 
\bibliographystyle{IEEEtran}

\end{document}